\documentclass[aip,apl,preprint]{revtex4-1}

\usepackage{graphicx}
\usepackage{dcolumn}
\usepackage{amsmath}
\usepackage[normalem]{ulem}

\makeatletter
\def\btt#1{\texttt{\@backslashchar#1}}
\DeclareRobustCommand\bblash{\btt{\@backslashchar}}
\makeatother

\usepackage{color}

\begin{document}

\preprint{HEP/123-qed}


\title[Short Title]{Circularly polarized cavity-mode emission from quantum dots\\
in a semiconductor three-dimensional chiral photonic crystal}

\author{S. Takahashi}
\affiliation{
Kyoto Institute of Technology, Sakyo, Kyoto 606-8585, Japan
}
\affiliation{
Graduate School of Engineering, Tohoku University, Aoba, Sendai, Miyagi 980-8579, Japan
}

\author{Y. Kinuta}
\affiliation{
Kyoto Institute of Technology, Sakyo, Kyoto 606-8585, Japan
}

\author{S. Ito}
\affiliation{
Kyoto Institute of Technology, Sakyo, Kyoto 606-8585, Japan
}

\author{H. Ohnishi}
\affiliation{
Kyoto Institute of Technology, Sakyo, Kyoto 606-8585, Japan
}

\author{K. Yamashita}
\affiliation{
Kyoto Institute of Technology, Sakyo, Kyoto 606-8585, Japan
}

\author{J. Tatebayashi}
\affiliation{
Institute for Nano Quantum Information Electronics, The University of Tokyo, Meguro, Tokyo 153-8505, Japan
}

\author{S. Iwamoto}
\affiliation{
Research Center for Advanced Science and Technology, The University of Tokyo, Meguro, Tokyo 153-8505, Japan
}
\affiliation{
Institute of Industrial Science, The University of Tokyo, Meguro, Tokyo 153-8505, Japan
}

\author{Y. Arakawa}
\affiliation{
Institute for Nano Quantum Information Electronics, The University of Tokyo, Meguro, Tokyo 153-8505, Japan
}


\date{\today}


\begin{abstract}
We experimentally demonstrated a circularly polarized cavity mode in a GaAs-based chiral photonic crystal (PhC) containing a planar defect. 
Low-temperature photoluminescence measurements of InAs quantum dots (QDs) embedded in the planar defect revealed a polarization bandgap for left-handed circularly polarized light in the near-infrared spectrum. 
Within this bandgap, where the QDs preferably emitted right-handed circularly polarized light, we observed a distinct cavity-mode peak characterized by left-handed circular polarization. 
This observation indicates that the chiral PhC modifies the optical density of states for left-handed circular polarization to be suppressed in the polarization bandgap and be largely enhanced at the cavity mode. 
The results obtained may not only provide photonic devices such as compact circularly polarized light sources but also promote strong coupling between circularly polarized photons and excitons in solid states or molecules, paving the way for advancements in polaritonics, spintronics, and quantum information technology. 
\end{abstract}



\maketitle


The chirality of circularly polarized light can be coupled with the chirality of matter. 
The spin angular momentum in a circularly polarized photon can be transferred to an electron/hole spin in solid states \cite{book}. 
This spin transfer has commonly been used in spintronics to polarize spin states or to generate circularly polarized light \cite{Ando,Iba}. 
In addition, the conversion of quantum states from a local spin qubit to a photon qubit via a spin-photon interface \cite{Greve,Gao} is crucial for quantum communication and memory. 
Circularly polarized light can also interact with chiral structures such as chiral molecules for sensing biochemical chirality and modifying chemical kinetics. 
To increase the coupling efficiency between circularly polarized light and spins or molecules, one of the simple approaches involves confining circularly polarized light in a small cavity that contains spins or chiral molecules. 
Since circular polarization (CP) is composed of orthogonal linear polarizations with $\pi$/2 phase difference, a cavity constructed from distributed Bragg reflectors whose top layer includes two-dimensional (2D) chiral structures has been demonstrated to confine circularly polarized light for a CP laser \cite{Sven}. 
However, such structures are mirror-symmetric, and thus do not provide CP as eigenpolarizations. 
Since the electric and magnetic fields of circularly polarized light exhibit a three-dimensional (3D) helical symmetry, mirror-asymmetric chiral or helical nanostructures offer CP eigenpolarization and efficient control of CP \cite{Gansel}.


One of the important characteristics of 3D helical nanostructures is polarization bandgaps in their photonic band structures. 
In a polarization bandgap, one type of CP light is predominantly reflected due to the helical modulation of refractive indices, whereas the other CP light is transmitted \cite{Lee,Vos}. 
In general, when a defect is introduced in a PhC that possesses a bandgap, defect modes or cavity modes appear in the bandgap \cite{Joanno-book,Sakoda-book}. 
This is also the case for polarization bandgaps \cite{review}. 
Cavities for circularly polarized light have been intensively studied in cholesteric liquid crystals, employing various methods to introduce defects in the helical structures, such as discontinuities in the helical phase \cite{PRL} or periodicity \cite{AdvMat}, as well as the incorporation of planar defects \cite{APL}. 
However, as the quantum spin transfer have been demonstrated in semiconductor self-assembled quantum dots (QDs) \cite{Greve,Gao} and a gate-defined semiconductor QDs \cite{Oiwa}, semiconductor-based nanostructures are required for applications in spintronic and quantum information technology. 
Additionally, semiconductor devices offer the advantage of monolithic compatibility with electrical circuits and can be driven by electrical current. 
In this study, we fabricated a GaAs-based chiral PhC with a planar defect containing InAs QDs as emitters in the near-infrared regime. 
The use of a planar defect allows for lots of gain materials in the 2D area towards the realization of a CP laser. 
Through photoluminescence (PL) measurements for the QDs in the absence of a magnetic field, we firstly observed a polarization bandgap for left-handed CP (LCP). 
In this bandgap, where QD emission was right-handed CP (RCP), we identified a cavity-mode peak exhibiting LCP, characterized by a quality factor of $Q \sim 390$. 
This results from the modification of the optical density of states for left-handed circular polarization, which was suppressed in the polarization bandgap and enhanced at the cavity mode. 
These results pave the way for the development of circularly polarized micro-lasers without external spin polarization, molecular chiral polaritons \cite{chiral_polariton}, and efficient spin convertors in spintronics and quantum information technology. 


\begin{figure}[t]
\includegraphics[width=0.6\linewidth]{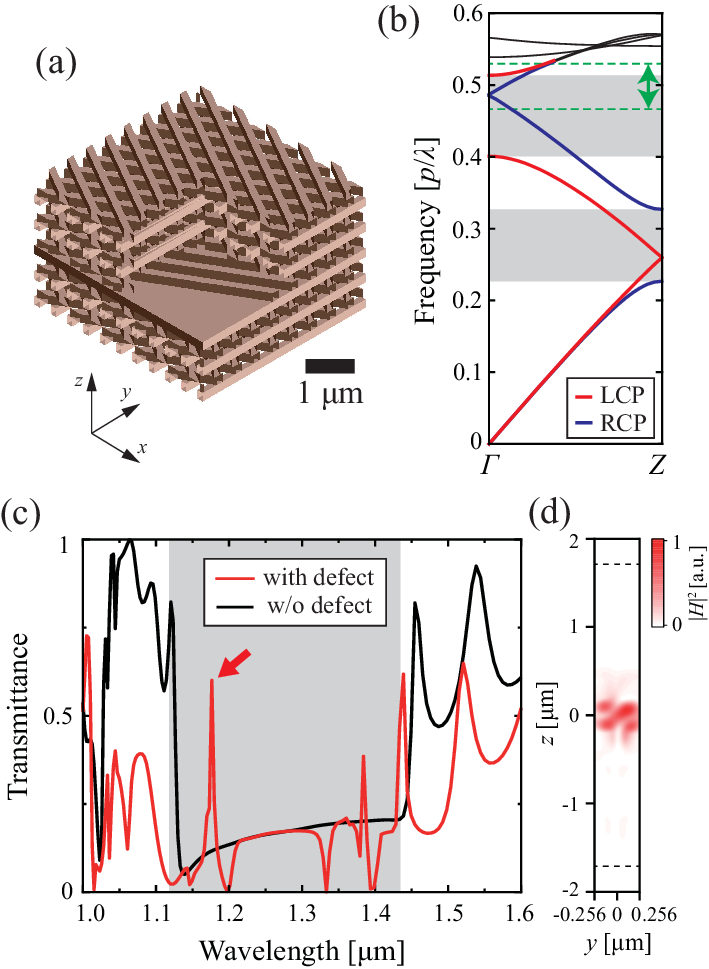}
\caption{\label{fig:structure}
(a)
Schematic diagram of the studied CP cavity. 
A planar defect is inserted in the chiral woodpile PhC. 
For clarity, the planar defect is exposed by cutting a part of the structure. 
(b)
Numerically calculated photonic band structure along the helical axis for the chiral PhC without the planar defect. 
The photonic band strongly polarized in LCP (RCP) is colored in red (blue).
The polarization band gaps for the opposite sense of CP appear in the two shaded regions. 
The frequency region indicated by the green double-headed arrow is measured in this study. 
(c)
Numerically obtained transmission spectra for the CP cavity with the defect and the chiral PhC without the defect. 
A peak of a CP cavity mode appears at a wavelength of 1176 nm in the shaded region of the polarization gap. 
(d)
Spatial distribution of the magnetic field at the CP cavity mode. 
The dotted lines indicate the edge of the structure. 
The intensity is localized at the planar defect as the central layer of the structure. 
}
\end{figure}


The studied 3D CP cavity is composed of a planar defect inserted in a chiral woodpile PhC, as shown in Fig. \ref{fig:structure}(a). 
The chiral PhC is constructed from a stack of thin plates having laterally periodic patterns of rods, each with a width of 150 nm, a period of 450 nm, and a thickness of 190 nm. 
The refractive index of the rods is 3.4, consistent with that of GaAs. 
These plates stacked one-by-one, with each plate rotated in-plane by 60$^{\circ}$ or 120$^{\circ}$. 
In this study, the chirality is right-handed with 120$^{\circ}$ rotation, wherein three plates form a single helical unit with a height of $p$ = 570 nm, as shown in Fig. \ref{fig:structure}(a) \cite{Vos}. 
For the chiral PhC without the defect, we calculated the photonic band structure by a plane wave expansion method. 
Figure \ref{fig:structure}(b) shows the resulting band structure in the stacking direction from $\it {\Gamma}$ to $Z$, which corresponds to the origin and edge of the first Brillouin zone in the $z$-direction, respectively. 
The optical modes represented by the red and blue bands are strongly polarized in LCP and RCP, respectively. 
Two polarization bandgaps appears at different frequencies, as indicated by the shaded regions in Fig. \ref{fig:structure}(b) \cite{Lee,TakahashiAPL,Vos}. 


In order to form CP cavity modes, we inserted a planar defect with a thickness of 380 nm and the same refractive index of 3.4 at the center of the chiral PhC. 
Figure \ref{fig:structure}(c) shows the LCP transmission spectra for the CP cavity (red curve) and the chiral PhC (black curve), calculated using the finite-difference time-domain (FDTD) method (See Appendix). 
The chiral PhC comprises 16 stacked plates, with the planar defect positioned centrally within the structure. 
The absence of the LCP modes in Fig. \ref{fig:structure}(b) results in low transmittance in the polarization bandgap (shaded region) for both spectra. 
In contrast, several transmission peaks appear in the red curve, indicating CP cavity modes. 
Figure \ref{fig:structure}(d) shows the spatial distribution of the magnetic field for one of the transmission peaks at a wavelength of 1176 nm, indicated by a red arrow in Fig. \ref{fig:structure}(c). 
As the dotted lines indicate the edge of the structure, the field intensity is localized around the planar defect, confirming a CP cavity mode. 
The quality-factor ($Q$-factor) of the CP cavity mode was evaluated to be $Q \sim$ 1167. 

\begin{figure}[t]
\includegraphics[width=0.8\linewidth]{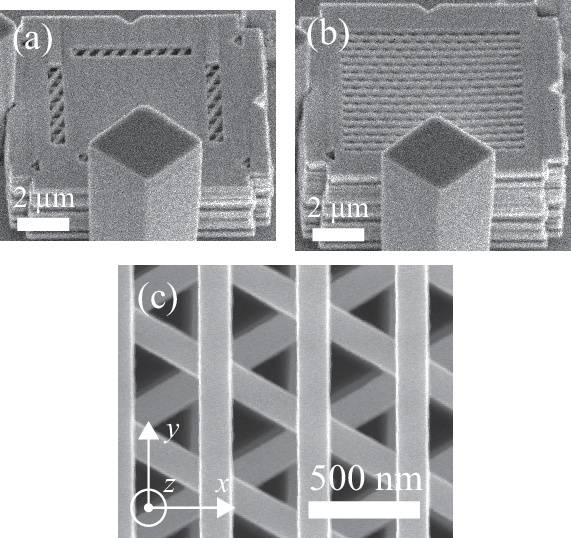}
\caption{\label{fig:sample}
(a), (b)
SEM images of the CP cavity in a perspective view, just after stacking the planar defect and the top plate, respectively. 
(c)
Magnified SEM image of the periodic rods in the top view. 
The crossing points arranged in a triangular lattice are aligned along the helical axis.
}
\end{figure}

Based on these numerical calculations, we fabricated the CP cavity with the same structural parameters as designed. 
Initially, we grew a sacrificial AlGaAs layer having a thickness of 1.5 $\mu$m on a GaAs substrate, followed by the growth of a 190 nm thick GaAs slab using metal-organic chemical vapor deposition. 
We then fabricated three types of 10 $\mu$m square plates having rod patterns rotated by 60$^{\circ}$ relative to one another \cite{TakahashiPRB}. 
These patterns were formed by electron beam lithography and subsequently transferred to the GaAs slab by reactive ion etching. 
Following this, the sacrificial layer was removed by wet etching to form suspended plates, which we referred to as passive plates. 
For the active plates, we prepared another substrate whose structure was the same as above, except that the GaAs slab contained InAs self-assembled QDs at a density of 10$^{10}$ /cm$^2$. 
For this substrate, we fabricated plates without any patterns, except for rectangular holes for the wet etching process, as shown in Fig. \ref{fig:sample}(a). 
Finally, these plates were stacked one-by-one using a micro-manipulation technique under optical microscope observation \cite{Kimura}. 
During the stacking process, we employed vertical posts to align the stacking plates. 
We stacked eight passive plates, followed by two active plates without lateral patterns, and then another eight passive plates on top. 
Here, in the absence of the two active plates working as a planar defect, the 16 passive plates form a chiral PhC with five helical periods. 
The 16th plate was stacked to be the same rod orientation as the first plate. 
Figure \ref{fig:sample}(a) and (b) show scanning electron microscope (SEM) images just after stacking the planar defect plate and the top plate, respectively. 
A magnified top view of the PhC is shown in Fig. \ref{fig:sample}(c). 
The stacking error was approximately 50 nm, which is an order of magnitude smaller than the structural period. 
Detailed fabrication methods for the plates and posts have been described in our previous reports \cite{Aoki2,Aniwat,TakahashiEL}. 


\begin{figure}[t]
\includegraphics[width=0.6\linewidth]{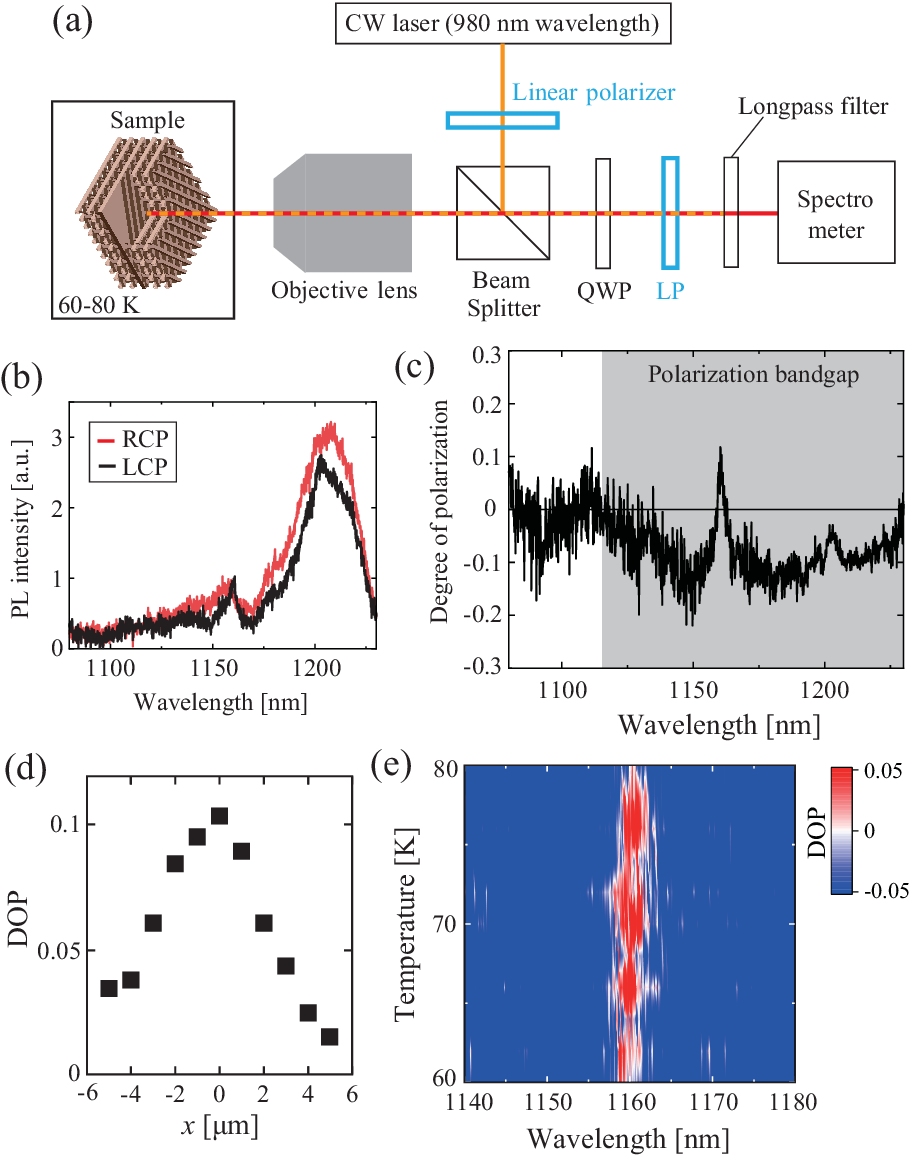}
\caption{\label{fig:PLmeasurement}
(a) Schematics diagram of the experimental setups for the PL measurement. 
(b) PL spectra for each CP component in the emission from QDs embedded in the defect of the chiral PhC at 66 K. 
(c) Spectrum of the degree of circular polarization obtained from (b). 
The shaded region indicates the polarization bandgap obtained from the numerical band structure in Fig. \ref{fig:structure}(b). 
The DOP is negative in this bandgap except for the peak at 1165 nm wavelength, indicating a CP cavity mode. 
(d) Position dependence of the peak DOP in Fig. (c) for the excitation laser spot. 
When the PhC center was displaced from the center of the laser spot, the peak DOP decreased, indicating that the CP cavity mode was localized at the chiral PhC.
(e) Color map of the DOP as a function of the emission wavelength and temperature. 
The temperature shift of the DOP peak was $\sim$0.05 nm/K, which is smaller than that of the bare QDs, confirming that the DOP peak was attributed to the CP cavity mode. 
}
\end{figure}


For this CP cavity, we performed photoluminescence (PL) measurements by optically exciting InAs wetting layers nearby the QDs in the defect. 
The measurement setup is schematically shown in Fig. \ref{fig:PLmeasurement}(a). 
As the excitation source, we used a continuous-wave semiconductor laser diode with a wavelength of 980 nm, which is outside the polarization bandgap and transparent to GaAs. 
The laser light with a power of 40 $\mu$W was linearly polarized, and focused onto the samples into a 5-$\mu$m spot size using a 50$\times$ objective lens with a numerical aperture of 0.65. 
Subsequently, the broadband infrared luminescence emitted from the QDs was collimated by the same objective lens and passed through a quarter-wave plate (QWP) and a linear polarizer (LP) to select the RCP and LCP components. 
The handedness-selected luminescence was finally detected by a spectrometer equipped with an array of InGaAs photodiodes, after eliminating the excitation laser component by a longpass filter. 
The sample was cooled down to 10-80 K, with its position finely tuned using piezoelectric nanopositioners. 
Note that no magnetic field was applied throughout the measurements. 


Figure \ref{fig:PLmeasurement}(b) shows the PL spectra of each CP component emitted from the CP cavity. 
In both spectra, two broad peaks appear around wavelengths of 1150 nm and 1210 nm, representing the ensemble QD emissions from the first excited state and the ground state, respectively (See Appendix). 
Comparing the two spectra, the emission intensity of LCP is slightly smaller than that of RCP around a wavelength of 1200 nm. 
To clearly compare these spectra, we calculated the degree of circular polarization, DOP = $(I_{LCP}-I_{RCP})/(I_{LCP}+I_{RCP})$, and plotted it in Fig. \ref{fig:PLmeasurement}(c) as a function of the emission wavelength. 
This figure distinctly shows that the DOP is negative, indicating that RCP light is preferably emitted from the QDs in the wavelength range from 1120 nm to 1230 nm. 
This occurs because the LCP bandgap suppresses the optical density of states of the vacuum field for the specific LCP electromagnetic mode depicted in Fig. ~\ref{fig:structure}(b) \cite{Konishi,TakahashiPRB}. 
The observed negative DOP region is consistent with the polarization bandgap obtained through numerical simulations presented in Fig. \ref{fig:structure}(b). 


In the polarization bandgap, a cavity mode peak was observed at a wavelength of 1160 nm, exhibiting a positive DOP of +0.1. 
The $Q$-factor for this CP cavity mode was evaluated as $Q \sim 390$. 
We measured the CP cavity mode as a function of the position of the laser spot. 
Initially, we aligned the center of the structure with the laser spot (zero displacement), then displaced the structure in the $x$-direction and measured the DOP. 
As shown in Fig. ~\ref{fig:PLmeasurement}(d), the DOP peak was spatially localized in the structure. 
In addition, we investigated the temperature dependence of the DOP spectrum from 60 K to 80 K. 
Figure ~\ref{fig:PLmeasurement}(e) shows a color map of the DOP spectra at various temperatures. 
The center wavelength of the DOP peaks, fitted using Lorentzian curves, exhibited a shift of $\sim$0.05 nm/K in the temperature range. 
In contrast, the bare QD emission showed a shift of $\sim$0.14 nm/K (See Appendix), indicating that the temperature sensitivity of the cavity mode is relatively lower. 
Based on these position-sensitive and temperature-insensitive results, we conclude that the DOP peak is attributed to a CP cavity mode \cite{AniwatAPL, TajiriAPL}. 
The measured CP cavity mode is consistent with the numerical results in Fig. ~\ref{fig:structure}(c). 
The slight differences in the peak wavelength and the $Q$-factor are probably caused by fabrication errors. 
Furthermore, we tested another CP cavity with slightly different structural parameters and found that a similar DOP peak in the polarization bandgap was reproducible (See Appendix). 


We note that the measured CP cavity mode exhibits a positive DOP. 
Since the RCP mode exists in the measured wavelength range and the QDs were excited using linearly polarized laser, the DOP at the cavity mode could be zero, when the CP cavity works as a simple CP bandpass filter. 
Consequently, the observed positive DOP signifies a substantial enhancement of the density of states of the vacuum field for the LCP mode in the vicinity of the planar defect \cite{TakahashiPRB}. 
In fact, our numerical calculations of radiative rates for LCP and RCP excitation at the planar defect indicate that the LCP radiative rates are larger than those for RCP around the cavity wavelength (See Appendix). 
This confirms that the density of states for the LCP mode is larger than that for the RCP mode. 


In summary, we have investigated a CP cavity mode in a 3D chiral PhC containing a planar defect, using photoluminescence measurements of InAs QDs in the defect. 
The emission spectra in the telecommunications wavelength regime revealed a LCP peak attributed to the cavity mode, which lies in a LCP polarization bandgap. 
The observed peak wavelength of 1160 nm and the $Q \sim$ 390 were in close agreement with the numerical calculations. 
The CP cavity in this semiconductor-based chiral PhC provides promising applications in various fields, such as photonics as a CP light source, in spintronics or quantum information technology as an efficient spin-photon interface, and in chemistry as molecular chiral polaritons. 


\begin{acknowledgments}

This work was supported by JST FOREST Program (Grant Number JPMJFR223Q, Japan). 

\end{acknowledgments}


\section*{Author Declarations}
The authors have no conflicts to disclose. 

\section*{Data Availability Statement}
The data that support the findings of this study are available from the corresponding author upon reasonable request. 


\appendix
\section{PL sprctra for bare InAs QDs}

\begin{figure}[t]
\includegraphics[width=1.0\linewidth]{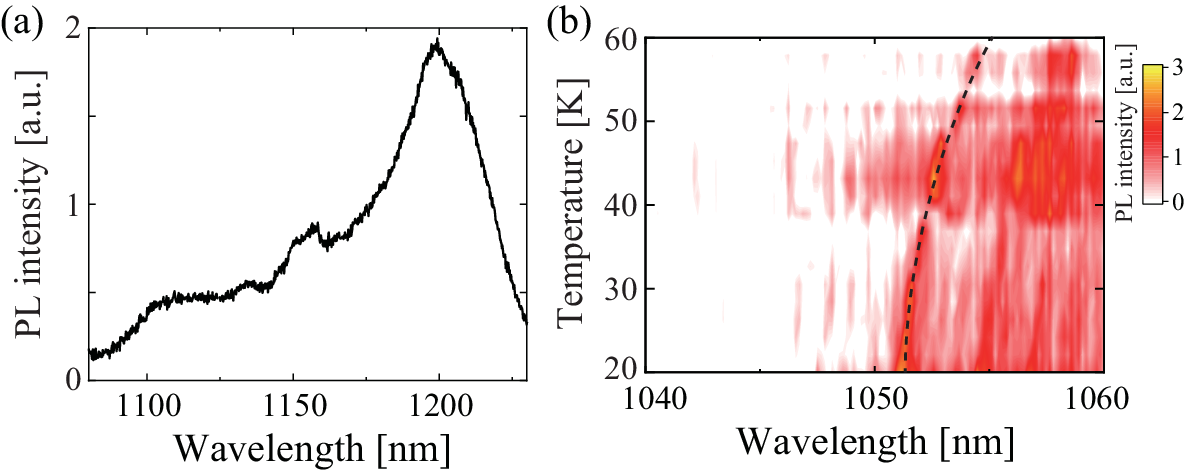}
\caption{\label{fig:supplement_QD}
(a) PL spectrum of the bare QDs at 60 K with an excitation power of 1 mW. 
(b) Color map of PL spectra for the bare QDs at various temperature. 
One of bright QDs shows an intensity peak at a wavelength of 1052.6 nm for 44 K with a temperature shift of 0.14 nm/K around 50 K, as indicated by a dotted line. 
}
\end{figure}

We measured PL spectra for the InAs QDs without any fabrication processes after the crystal growth. 
Figure ~\ref{fig:supplement_QD}(a) shows a PL spectrum for the bare QDs at 60 K. 
The excitation laser light with a power of 1 mW was linearly polarized. 
Two broad peaks appear around wavelengths of 1150 nm and 1210 nm, representing the ensemble QD emissions from the first excited state and the ground state, respectively. 
The PL intensity in a wavelength range of 1100-1140 nm is probably caused by higher excited states. 
Figure ~\ref{fig:supplement_QD}(b) shows PL spectra of the bare QDs at various temperature. 
One of slightly bright QDs shows an intensity peak in ensemble QD emissions at a wavelength of 1052.6 nm for 44 K. 
This peak wavelength is shifted by temperature to be 1054.6 nm at 58 K. 
This temperature shift is 0.14 nm/K in the assumption of linear dependence on temperature around 50 K, whereas the shift is not proportional to temperature around 30 K. 


\section{Experimental results on another CP cavity}

\begin{figure}[t]
\includegraphics[width=1.0\linewidth]{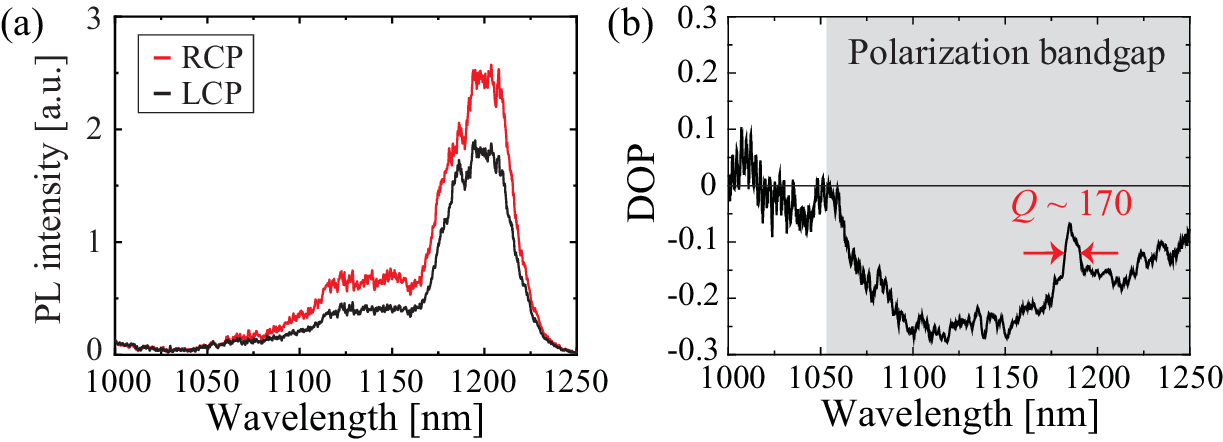}
\caption{\label{fig:supplement_exp}
(a) PL spectra for each CP component in the emission from QDs embedded in the planar defect. 
The PL measurement was performed at 15 K. 
(b) Spectrum of the degree of circular polarization obtained from (a). 
}
\end{figure}

We fabricated another CP cavity by using micromanipulation technique under scanning electron microscope observation \cite{Aoki2,Aniwat}. 
The structural parameters were slightly different from the CP cavity in the main text. 
About the in-plane pattern, the width and thickness of the rods were 160 nm and 200 nm, and the period was the same. 
The thickness of the planar defect was 400 nm by stacking two active plates without the in-plane pattern. 
The measured emission spectra for each CP component are shown in Fig. ~\ref{fig:supplement_exp}(a). 
From these two spectra, the DOP at each wavelength was plotted, as shown in Fig. ~\ref{fig:supplement_exp}(b). 
In the negative DOP region corresponding to the polarization band gap, we can find a small peak at a wavelength of 1185 nm. 
Though the DOP at the peak top is negative, the wavelength and $Q \sim$ 170 of the cavity-mode peak are consistent with numerical results. 


\section{Details of the numerical calculations}

The numerical results in Fig. ~\ref{fig:structure}(c) was obtained by using a FDTD method. 
We performed two independent numerical simulations for different CP incidence to each 3D structure. 
The calculation region was defined by periodic boundary conditions in the $x$ and $y$ directions and perfectly matched layers attached at 10 $\mu$m away from the center of the structure in the $z$ direction. 
A pulsed plane wave with LCP was irradiated in the $z$ direction. 
The pulse duration was a single cycle for the central wavelength of 1250 nm. 
The recorded time-dependent electromagnetic field of the transmitted light was analyzed in frequency space by a Fourier transform. 


For the calculations of spontaneous emission rates, a classical dipole radiation power was investigated by the FDTD method \cite{Xu-Yariv,TakahashiPRB}. 
The numerically studied CP cavity has the same dimensions as the experimental structure except for the in-plane area of the rod pattern, owing to computational time restrictions. 
The area for the calculation was a 3.6 $\mu$m square corresponding to 8 in-plane periods, compared to that in the experiment was the 7.2 $\mu$m square corresponding to 16 in-plane periods. 
As the measured structure, this in-plane area was numerically surrounded in the in-plane directions by GaAs with a refractive index of 3.4 which is same as the rods.
We set perfectly matched layers at 2 periods (pitches) away from the edge of the pattern in the in-plane ($z$) directions. 
Also, we set spatial domain as 1/16 period (1/48 pitch) in the in-plane ($z$) directions. 

\begin{figure}[t]
\includegraphics[width=1.0\linewidth]{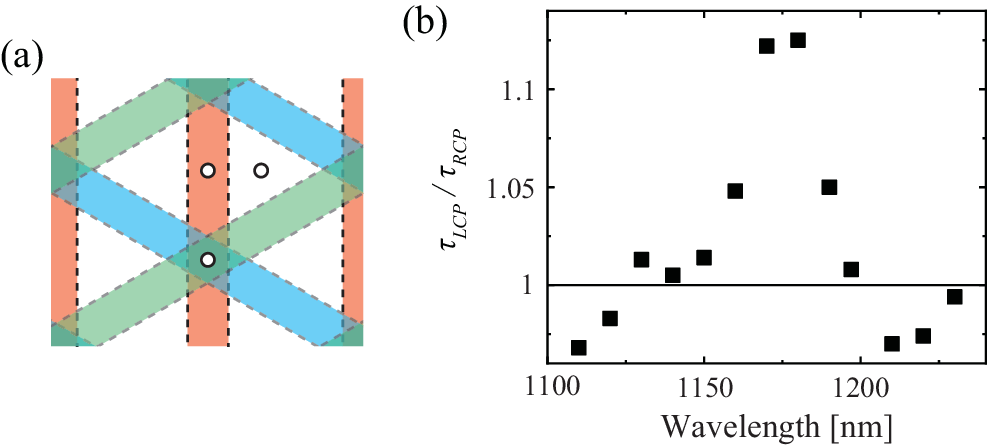}
\caption{\label{fig:supplement_cal}
(a) Schematic top view of the CP cavity. 
The rods in each plate are colored differently. 
The white circles indicate the positions of the point source in the planar defect. 
(b) Calculated lifetime ratio $\tau_{LCP}/\tau_{RCP}$ as a function of wavelength. 
}
\end{figure}

Then we put a current source with each CP at particular positions. 
Figure ~\ref{fig:supplement_cal}(a) shows a schematic top view of the CP cavity . 
Owing to computational time restrictions, we put light sources at six points in the planar defect, though quantum dots in the experiment were contained in the rods randomly with high density. 
The three out of six positions are indicated by white circles in Fig. ~\ref{fig:supplement_cal}(a), and since the planar defect was composed of two plates in the experiment, we put the source at the central height of each defect plate, thus the total position of the source is six. 
The size of the light sources is the same as the grid size to be a point source. 
The electromagnetic field is generated by a circular dipole in the $x$-$y$ plane, because the studied QDs are almost flat with a $\sim$ 10 nm diameter and a $\sim$ 3 nm height. 


In these conditions, we calculate the Poynting vectors at a surface which is set at one period/pitch away from the edge of the pattern and encloses the light source, then integrate the obtained Poynting vectors over the whole surface and the whole evolution time. 
We performed this numerical method for 6 different source positions, 13 different emission wavelengths, and 2 orthogonal circular polarization with and without the chiral PhC, totally 312 simulations. 
Then the integrated power for the CP cavity were normalized by the power without the structure, and we obtained the modification coefficients of the spontaneous emission rate. 
The modification coefficients are finally averaged for the 6 simulations with different source positions. 
Figure ~\ref{fig:supplement_cal}(b) shows the lifetime ratio $\tau_{LCP}/\tau_{RCP}$ as a function of emission wavelength. 
From this figure, we can find a braod peak of $\tau_{LCP}/\tau_{RCP} >$ 1 around a wavelength of 1170 nm. 
This peak wavelength is consistent with that in the transmission spectrum in Fig. ~\ref{fig:structure}(c) (1176 nm). 
Also, $\tau_{LCP}/\tau_{RCP} <$ 1 around 1120 nm and 1210 nm is consistent with the LCP polarization band gap. 
The broad linewidth was probably caused by the small in-plane size of the calculations, owing to computational time restrictions. 


\begin{thebibliography}{00}

\bibitem{book} F. Meier and B. P. Zakharchenya, {\it Optical Orientation} (North-Holland, Amsterdam, 2012).

\bibitem{Ando}
H. Ando, T. Sogawa, and H. Gotoh, ``Photon-spin controlled lasing oscillation in surface-emitting lasers,'' Appl. Phys. Lett. {\bf 73}, 566 (1998).

\bibitem{Iba}
S. Iba, S. Koh, K. Ikeda, and H. Kawaguchi, ``Room temperature circularly polarized lasing in an optically spin injected vertical-cavity surface-emitting laser with (110) GaAs quantum wells,'' Appl. Phys. Lett. {\bf 98}, 081113 (2011).

\bibitem{Greve} K. D. Greve, L. Yu, P. L. McMahon, J. S. Pelc, C. M. Natarajan, N. Y. Kim, E. Abe, S. Maier, C. Schneider, M. Kamp, S. Hofling, R. H. Hadfield, A. Forchel, M. M. Fejer, and Y. Yamamoto, ``Quantum-dot spin-photon entanglement via frequency downconversion to telecom wavelength,'' Nature {\bf 491,} 421--425 (2012).

\bibitem{Gao} W. B. Gao, P. Fallahi, E. Togan, J. M. Sanchez, and A. Imamoglu, ``Observation of entanglement between a quantum dot spin and a single photon,'' Nature {\bf 491,} 426--429 (2012).

\bibitem{Sven}
A. A. Demenev, V. D. Kulakovskii, C. Schneider, S. Brodbeck, M. Kamp, S. H$\rm {\ddot{o}}$fling, S. V. Lobanov, T. Weiss, N. A. Gippius, and S. G. Tikhodeev, ``Circularly polarized lasing in chiral modulated semiconductor microcavity with GaAs quantum wells,'' Appl. Phys. Lett. {\bf 109}, 171106 (2016).

\bibitem{Gansel} J. K. Gansel, M. Thiel, M. S. Rill, M. Decker, K. Bade, V. Saile, G. von Freymann, S. Linden, and M. Wegener, ``Gold helix photonic metamaterial as braodband circular polarizer,'' Science {\bf 325,} 1513--1515 (2009).

\bibitem{Lee} J. C. W. Lee and C. T. Chan, ``Polarization gaps in spiral photonic crystals,'' Optics Express {\bf 13,} 8083--8088 (2005).

\bibitem{Vos} S. Takahashi, T. Tajiri, Y. Arakawa, S. Iwamoto, and W. L. Vos, ``Optical properties of chiral three-dimensional photonic crystals,'' Phys, Rev, B {\bf 107}, 165307 (2023).

\bibitem{Joanno-book} J. D. Joannopoulos, S. G. Johnson, J. N. Winn, and R. D. Meade, {\it Photonic Crystals: Molding the Flow of Light, \rm{2nd ed.}} (Princeton University Press, Princeton NJ, 2008).

\bibitem{Sakoda-book} K. Sakoda, {\it Optical Properties of Photonic Crystals, \rm{2nd ed.}} (Springer, New York, 2005).

\bibitem{review} V. I. Kopp, Z.-Q. Zhang, and A. Z. Genack, ``Lasing in chiral photonic structures,'' Prog. Quantum Electron. {\bf 27,} 369--416 (2003).

\bibitem{PRL} J. Schmidtke, W. Stille, and H. Finkelmann, ``Defect Mode Emission of a Dye Doped Cholesteric Polymer Network,'' Phys. Rev. Lett. {\bf 90,} 083902 (2003).

\bibitem{AdvMat} M. H. Song, N. Y. Ha, K. Amemiya, B. Park, Y. Takanishi, K. Ishikawa, J. W. Wu, S. Nishimura, T. Toyooka, H. Takezoe, ``Defect-Mode Lasing with Lowered Threshold in a Three-Layered Hetero-Cholesteric Liquid-Crystal Structure,'' Adv. Matter. {\bf 18,} 193--197 (2006).

\bibitem{APL} S. M. Jeong, N. Y. Ha, Y. Takanishi, K. Ishikawa, H. Takezoe, S. Nishimura, and G. Suzaki, ``Defect mode lasing from a double-layered dye-doped polymeric cholesteric liquid crystal films with a thin rubbed defect layer,'' Appl. Phys. Lett. {\bf 90}, 261108 (2007).

\bibitem{Oiwa} T. Fujita, K. Morimoto, H. Kiyama, G. Allison, M. Larsson, A. Ludwig, S. R. Valentin, A. D. Wieck, A. Oiwa, and S. Tarucha, ``Angular momentum transfer from photon polarization to an electron spin in a gate-defined quantum dot,'' Nat. Commun. {\bf 10,} 2991 (2019).

\bibitem{chiral_polariton} D. G. Baranov, C. Sch$\rm{\ddot{a}}$fer, and M. V. Gorkunov, ``Toward Molecular Chiral Polaritons,'' ACS Photonics {\bf 10,} 2440--2455 (2023).

\bibitem{TakahashiAPL}
S. Takahashi, T. Tajiri, Y. Ota, J. Tatebayashi, S. Iwamoto, and Y. Arakawa, Appl. Phys. Lett. {\bf 105}, 051107 (2014).

\bibitem{Kimura} S. Takahashi, E. Kimura, T. Ishida, T. Tajiri, K. Watanabe, K. Yamashita, S. Iwamoto, and Y. Arakawa, ``Fabrication of three-dimensional photonic crystals for near-infrared light by micro-manipulation technique under optical microscope observation,'' Appl. Phys. Express {\bf 15}, 015001 (2021).

\bibitem{Aoki2} K. Aoki, D. Guimard, M. Nishioka, M. Nomura, S. Iwamoto, and Y. Arakawa, ``Coupling of quantum-dot light emission with a three-dimensional photonic-crystal nanocavity,'' Nat. Photonics {\bf 2,} 688--692 (2008).

\bibitem{Aniwat} A. Tandaechanurat, S. Ishida, D. Guimard, M. Nomura, S. Iwamoto, and Y. Arakawa, ``Lasing oscillation in a three-dimensional photonic crystal nanocavity with a complete bandgap,'' Nat. Photonics {\bf 5,} 91--94 (2010).

\bibitem{TakahashiEL} S. Takahashi, T. Tajiri, K. Watanabe, Y. Ota, S. Iwamoto, and Y. Arakawa, ``High-\textit{Q} nanocavities in semiconductor-based three-dimensional photonic crystals,'' Electron. Lett. {\bf 54}, 305 (2018).

\bibitem{Konishi}
K. Konishi, M. Nomura, N. Kumagai, S. Iwamoto, Y. Arakawa, and M. K.-Gonokami, ``Circularly polarized light emission from semiconductor planar chiral nanostructures,'' Phys. Rev. Lett. {\bf 106,} 057402 (2011).

\bibitem{TakahashiPRB}
S. Takahashi, Y. Ota, T. Tajiri, J. Tatebayashi, S. Iwamoto, and Y. Arakawa, ``Circularly polarized vacuum field in three-dimensional chiral photonic crystals probed by quantum dot emission,'' Phys. Rev. B {\bf 96}, 195404 (2017).

\bibitem{AniwatAPL} A. Tandaechanurat, S. Ishida, K. Aoki, D. Guimard, M. Nomura, S. Iwamoto, and Y. Arakawa, ``Demonstration of high-$Q$ ($>$8600) three-dimensional photonic crystal nanocavity embedding quantum dots,'' Appl. Phys. Lett. {\bf 94}, 171115 (2009).

\bibitem{TajiriAPL} T. Tajiri, S. Takahashi, Y. Ota, J. Tatebayashi, S. Iwamoto, and Y. Arakawa, ``Demonstration of a three-dimensional photonic crystal nanocavity in a $<$110$>$-layered diamond structure,'' Appl. Phys. Lett. {\bf 107}, 071102 (2015).

\bibitem{Xu-Yariv} Y. Xu, {\it et al.}, J. Opt. Soc. Am. B {\bf 16}, 465 (1999).
\end{thebibliography}
\end{document}